\documentclass[12pt]{article}

\usepackage{graphicx}
\usepackage{epsfig}
\usepackage{color}

\def\gtwid{\mathrel{\raise.3ex\hbox{$>$\kern-.75em\lower1ex\hbox{$\sim$}}}}
\def\ltwid{\mathrel{\raise.3ex\hbox{$<$\kern-.75em\lower1ex\hbox{$\sim$}}}}
\def\square{\kern1pt\vbox{\hrule height 1.2pt\hbox{\vrule width 1.2pt\hskip 3pt
   \vbox{\vskip 6pt}\hskip 3pt\vrule width 0.6pt}\hrule height 0.6pt}\kern1pt}

\begin{document}

\begin{titlepage}

\begin{flushright}
UFIFT-QG-17-03 \\
\end{flushright}

\vskip 1cm

\begin{center}
{\bf Deducing Cosmological Observables from the S-matrix}
\end{center}

\vskip .5cm

\begin{center}
S. P. Miao$^{1*}$, T. Prokopec$^{2\star}$ and R. P. Woodard$^{3\dagger}$
\end{center}

\vskip .5cm

\begin{center}
\it{$^{1}$ Department of Physics, National Cheng Kung University \\
No. 1, University Road, Tainan City 70101, TAIWAN}
\end{center}

\begin{center}
\it{$^{2}$ Institute for Theoretical Physics, Spinoza Institute \& EMME$\Phi$ \\
Utrecht University, Postbus 80.195, 3508 TD Utrecht, THE NETHERLANDS}
\end{center}

\begin{center}
\it{$^{3}$ Department of Physics, University of Florida,\\
Gainesville, FL 32611, UNITED STATES}
\end{center}

\vspace{.5cm}

\begin{center}
ABSTRACT
\end{center}

We study one loop quantum gravitational corrections to the long range 
force induced by the exchange of a massless scalar between two massive
scalars. The various diagrams contributing to the flat space S-matrix 
are evaluated in a general covariant gauge and we show that dependence 
on the gauge parameters cancels at a point considerably {\it before} 
forming the full S-matrix, which is unobservable in cosmology. It is 
possible to interpret our computation as a solution to the effective 
field equations --- which could be done even in cosmology --- but 
taking account of quantum gravitational corrections from the source 
and from the observer. 

\begin{flushleft}
\end{flushleft}

\vskip .5cm

\begin{flushleft}
$^{*}$ e-mail: spmiao5@mail.ncku.edu.tw \\
$^{\star}$ e-mail: T.Prokopec@uu.nl \\
$^{\dagger}$ e-mail: woodard@phys.ufl.edu
\end{flushleft}

\end{titlepage}

\section{Introduction} \label{intro}

Primordial inflation produces a vast ensemble of long wave length gravitons,
which is what causes the tensor power spectrum \cite{Starobinsky:1979ty}. It
is inconceivable that these gravitons simply exist, without interacting, at
some level, with themselves and other particles. If such an ensemble were
present today no one doubts that it would change the way particles propagate, 
or that it might affect the long range forces carried by virtual particles. 
Indeed, the effect of gravitational radiation on the propagation of photons 
is the basis for using pulsar timing to detect gravitational radiation 
\cite{Detweiler:1979wn,Lorimer:2008se}.

Although the actual geometry of inflation must show evolution, for many
purposes one can employ the simpler, de Sitter geometry as a reasonable 
approximation. The effects of inflationary gravitons on a particle's 
kinematics, and on the force it carries, are studied in the same way. One
first computes the one graviton loop correction to the particle's 1PI 
(one-particle-irreducible) 2-point function. Then one uses this result to
quantum-correct the linearized effective field equation for the particle.
Many effects have been studied in this way over the course of the past
decade:
\begin{itemize}
\item{Inflationary gravitons induce a progressive excitation of massless 
and light fermions which eventually becomes nonperturbatively strong 
\cite{Miao:2005am,Miao:2006gj,Miao:2007az,Miao:2012bj} due to the spin-spin 
coupling \cite{Miao:2008sp};}
\item{Inflationary gravitons have little effect on massless, minimally 
coupled scalars \cite{Kahya:2007bc,Kahya:2007cm} owing to the absence of
such a coupling;}
\item{Inflationary gravitons secularly excite photons as they do fermions 
\cite{Leonard:2013xsa,Wang:2014tza,Glavan:2015ura,Glavan:2016bvp}, and 
they also induce secular modifications of electrodynamic forces
\cite{Glavan:2013jca};}
\item{Inflationary gravitons secularly excite other gravitons 
\cite{Tsamis:1996qk,Mora:2013ypa}; and}
\item{Inflationary gravitons secularly excite conformally coupled scalars
owing to the conformal coupling \cite{Boran:2014xpa,Boran:2017fsx,
Boran:2017cfj}.}
\end{itemize}

The physics behind these results seems plausible enough: inflationary 
gravitons scatter particles and force carriers, with the net 
deviation growing as the particle or force carrier propagates further. 
However, the reality of these effects is thrown into question by the 
notorious {\it gauge issue}. The graviton propagator depends upon an 
arbitrary gauge choice, which certainly affects full 1PI functions on 
flat space background. On the other hand, certain parts of the flat
1PI $N$-point functions are gauge independent because sums of products 
of them combine to form the gauge independent S-matrix. So it seemed
possible that the leading secular dependence on de Sitter background
might be independent of the gauge \cite{Miao:2012xc}.

Computations on de Sitter background are so terribly difficult that 
almost all work has been done in a single, particularly simple gauge \cite{Tsamis:1992xa,Woodard:2004ut}. However, a determined effort at
length produced a result for the graviton correction to the vacuum 
polarization \cite{Glavan:2015ura} in a one-parameter family of de 
Sitter invariant gauges \cite{Mora:2012zi}. When this was used to
quantum-correct Maxwell's equation the result was that dynamical 
photons suffer a progressive excitation which is independent of the
gauge parameter \cite{Glavan:2016bvp}. The excitation has the same sign
and time dependence as for the simple gauge \cite{Wang:2014tza}, but 
the numerical coefficient is not quite the same. Hence it seems that
even the leading secular effects of inflationary gravitons are somewhat
gauge dependent.

On flat space background this sort of issue would be resolved by 
reference to the S-matrix, which is gauge independent. Unfortunately,
that option is not available for inflationary cosmology. A synthetic S-matrix 
has been shown to exist for massive fields on de Sitter \cite{Marolf:2012kh},
but causality precludes an inflationary observer from making the global
measurements it requires. 

Another alternative would be to devise gauge invariant operators to quantify
changes in particle kinematics and force laws, then compute the expectation
values of these operators. One major disadvantage of this technique is
that there are no local invariants in gravity, so any observables would be
nonlocal composite operators. That vastly complicates renormalization. It 
is just possible to persevere through such a computation \cite{Miao:2017vly},
but it seems worth looking for a simpler technique. It would be particularly 
nice to devise some way of modifying the effective field equations.

A promising approach is the one developed by Donoghue \cite{Donoghue:1993eb,
Donoghue:1994dn}, who noted that the leading quantum gravitational corrections
to long range forces derive from a very special sort of nonanalytic correction
to loop amplitudes. This is typically implemented by computing scattering
amplitudes in Fourier momentum space, extracting the important contribution to 
the S-matrix, and then inferring corrections to the potential by inverse 
scattering. For example, this is how the first complete one loop computations 
were made of quantum gravitational corrections to the Newtonian potential
\cite{BjerrumBohr:2002ks,BjerrumBohr:2002kt}. However, we suspect that the 
essential part of the technique can be separated from the full S-matrix, and 
phrased instead as a way of correcting the effective field equations --- in
position space --- to include quantum gravitational correlations with the 
source which disturbs the effective field and the observer who measures it.

To examine this possibility we have chosen to work on flat space background, 
with the object of computing the one graviton loop correction to the long 
range potential induced by a massless scalar. In section~\ref{want} we make
the computation in the manner described above: calculating the scalar self-mass,
then using it to quantum correct the effective field equation. By making the
computation in the 2-parameter family of Poincar\'e invariant gauges we 
demonstrate that the result is highly gauge dependent. In section~\ref{bryce} 
we review an intriguing comment of the subject made by one of the great men of 
quantum gravity. In section~\ref{obssource} we abstract Donoghue's technique to 
position space, and show how it can be viewed as correcting the effective field 
equation. We explicitly demonstrate that the gauge dependence cancels when all
the corrections are included. Section~\ref{discuss} summarizes what we have 
shown and discusses the implications for cosmology.

In addition to the obvious debt we owe to Donoghue, it should be noted that 
section~\ref{obssource} closely follows Bjerrum-Bohr's computation of the 
quantum gravitational correction to the Coulomb potential in SQED (scalar 
quantum electrodynamics) \cite{BjerrumBohr:2002sx}. The relevant diagram 
topologies are the same once one replaces his photon lines with our massless
scalar lines. Also, we merely translated to position space the three
Fourier momentum integrals he used to extract the leading infrared 
contributions.

\section{What We Wanted} \label{want}

In this section we discuss a simple, flat space analog of the sort of computations
we have been doing of how inflationary gravitons change particle kinematics and 
force laws. The quantity we have chosen to correct is the long range force exerted
by a massless scalar. We begin by reviewing the Feynman rules for the most general
Poncar\'e invariant gauge. Then the one graviton loop contribution to the scalar
self-mass is computed in dimensional regularization and fully renormalized. Finally,
we use this result to quantum-correct the scalar field equation, and we solve for
the response to a static point source. 

\subsection{Feynman Rules} \label{feynman}

The Lagrangian of gravity plus a massless, minimally coupled scalar is,
\begin{equation}
\mathcal{L}_1 = \frac{R \sqrt{-g}}{16 \pi G} -\frac12 \partial_{\mu} \phi 
\partial_{\nu} \phi g^{\mu\nu} \sqrt{-g} \; . \label{Lag1}
\end{equation}
We are perturbing around flat space with the usual definitions of the graviton 
field $h_{\mu\nu}$ and the loop counting parameter $\kappa^2$,
\begin{equation}
g_{\mu\nu}(x) \equiv \eta_{\mu\nu} + \kappa h_{\mu\nu}(x) \qquad , \qquad
\kappa^2 \equiv 16 \pi G \; . \label{graviton}
\end{equation}
By convention graviton indices are raised and lowered using the Lorentz metric,
$h^{\mu\nu} \equiv \eta^{\mu\rho} \eta^{\nu\sigma} h_{\rho\sigma}$, $h \equiv
\eta^{\mu\nu} h_{\mu\nu}$. The expansions we require are, 
\begin{equation}
\sqrt{-g} = 1 + \frac12 \kappa h + \dots \qquad , \qquad g^{\mu\nu} = 
\eta^{\mu\nu} - \kappa h^{\mu\nu} + \dots \label{expansion} 
\end{equation}
When using dimensional regularization on flat space background the order
$\kappa^2$ interactions are not necessary for the sorts of diagrams we 
require.

To facilitate dimensional regularization we work in $D$-dimensional spacetime.
Because temporal Fourier transforms are problematic in cosmology, we make this
calculation in position space. The massless scalar propagator is,
\begin{equation}
i\Delta(x;x') = \frac{\Gamma(\frac{D}2 \!-\! 1)}{4 \pi^{\frac{D}2} \Delta 
x^{D-2}} \quad , \quad \Delta x^{\mu} \equiv (x \!-\! x')^{\mu} \quad , \quad
\Delta x^2 \equiv \eta_{\mu\nu} \Delta x^{\mu} \Delta x^{\nu} \; . 
\label{scalarprop}
\end{equation}
The scalar propagator obeys an equation of great significance for us,
\begin{equation}
\partial^2 i\Delta(x;x') = i\delta^D(x \!-\! x') \; . \label{propeqn}
\end{equation}

We fix the gauge by adding to the Lagrangian the most general Poincar\'e 
invariant gauge fixing term,
\begin{equation}
\mathcal{L}_{GF} = -\frac1{2a} \eta^{\mu\nu} \mathcal{F}_{\mu} 
\mathcal{F}_{\nu} \qquad , \qquad \mathcal{F}_{\mu} \equiv \eta^{\rho\sigma}
\Bigl( h_{\mu \rho , \sigma} \!-\! \frac{b}{2} h_{\rho\sigma , \mu}\Bigr)
\; . \label{gauge}
\end{equation}
The resulting graviton propagator can be expressed in terms of the massless
scalar propagator using the transverse projection operator $\Pi_{\mu\nu} 
\equiv \eta_{\mu\nu} - \partial_{\mu} \partial_{\nu}/\partial^2$ as 
\cite{Capper:1979ej},
\begin{eqnarray}
\lefteqn{ i\Bigl[\mbox{}_{\mu\nu} \Delta_{\rho\sigma}\Bigr](x;x') = \Biggl\{
2 \Pi_{\mu (\rho} \Pi_{\sigma )\nu} - \frac{2}{D \!-\! 1} \Pi_{\mu\nu}
\Pi_{\rho\sigma} } \nonumber \\
& & \hspace{1cm} -\frac{2}{(D \!-\! 2) (D \!-\! 1)} \Biggl[ \eta_{\mu\nu} -
\Bigl( \frac{D b \!-\! 2}{b \!-\! 2}\Bigr) \frac{\partial_{\mu} \partial_{\nu}}{
\partial^2} \Biggr] \Biggl[ \eta_{\rho\sigma} - \Bigl( \frac{D b \!-\! 2}{b \!-\! 
2}\Bigr) \frac{\partial_{\rho} \partial_{\sigma}}{\partial^2} \Biggr] \nonumber \\
& & \hspace{2cm} + 4 a \!\times\! \frac{\partial_{(\mu} \Pi_{\nu ) (\rho} 
\partial_{\sigma)}}{\partial^2} + \frac{4a}{(b \!-\! 2)^2} \!\times\!
\frac{\partial_{\mu} \partial_{\nu} \partial_{\rho} \partial_{\sigma}}{\partial^4}
\Biggr\} i\Delta(x;x') \; . \qquad \label{gravprop}
\end{eqnarray}
Here and henceforth parenthesized indices are symmetrized. The factors of 
$\partial_{\mu} \partial_{\nu}/\partial^2$ acting on the massless scalar
propagator $i\Delta(x;x')$ can be written as \cite{Leonard:2012fs},
\begin{eqnarray}
\frac{\partial_{\mu} \partial_{\nu}}{\partial^2} i\Delta(x;x') & = & \frac12 
\!\times\! \Biggl\{\eta_{\mu\nu} \!-\! \frac{(D \!-\! 2) \Delta x_{\mu} 
\Delta x_{\nu}}{\Delta x^2} \Biggr\} i\Delta(x;x') \; , \\
\frac{\partial_{\mu} \partial_{\nu} \partial_{\rho} \partial_{\sigma}}{\partial^4} 
i\Delta(x;x') & = & \frac18 \!\times\! \Biggl\{ 3 \eta_{(\mu\nu} \eta_{\rho\sigma)} 
\!-\! \frac{6 (D \!-\! 2) \eta_{(\mu\nu} \Delta x_{\rho} 
\Delta x_{\sigma)}}{\Delta x^2} \nonumber \\
& & \hspace{1cm} + \frac{D (D \!-\! 2) \Delta x_{\mu} \Delta x_{\nu} 
\Delta x_{\rho} \Delta x_{\sigma}}{\Delta x^4} \Biggr\} i\Delta(x;x') \; . \qquad
\end{eqnarray}

\subsection{One Loop Self-Mass} \label{selfmass}

The scalar self-mass $-i M^2(x;x')$ is the 1PI (1-particle irreducible) scalar
2-point function. The primitive one graviton loop correction to it is,
\begin{eqnarray}
\lefteqn{ -i M^2(x;x') = \partial_{\alpha} \partial'_{\beta} \Biggl\{
i\Bigl[\mbox{}_{\mu\nu} \Delta_{\rho\sigma}\Bigr](x;x') \!\times\! (-i\kappa)
\Bigl[-\eta^{\alpha\mu} \partial^{\nu} \!+\! \frac12 \eta^{\mu\nu} 
\partial^{\alpha}\Bigr] } \nonumber \\
& & \hspace{5cm} \times (-i\kappa) \Bigl[ -\eta^{\beta\rho} {\partial'}^{\sigma}
\!+\! \frac12 \eta^{\rho\sigma} {\partial'}^{\beta}\Bigr] i\Delta(x;x') \Biggr\} .
\qquad \label{primitiveM}
\end{eqnarray}
After performing the tensor contractions and making use of the identities,
\begin{eqnarray}
\frac1{\Delta x^{2D-2}} & = & \frac{\partial^2}{2 (D \!-\! 2)^2} 
\frac1{\Delta x^{2D-4}} \; , \\
\frac{\Delta x^{\mu} \Delta x^{\nu}}{\Delta x^{2D}} & = & \Biggl[
\frac{\eta^{\mu\nu} \partial^2}{4 (D\!-\!2)^2 (D\!-\!1)} + \frac{\partial^{\mu}
\partial^{\nu}}{4 (D \!-\! 2) (D \!-\! 1)} \Biggr] \frac1{\Delta x^{2D-4}} 
\; . \qquad 
\end{eqnarray}
we reach an expression in terms of the square of scalar propagator (\ref{scalarprop}),
\begin{equation}
-i M^2(x;x') = -\kappa^2 \mathcal{C}_0(a,b,D) \!\times\! \partial^4 \Bigl[ i
\Delta(x;x')\Bigr]^2 \; . \label{simpleM}
\end{equation}
The gauge dependence resides in the multiplicative factor,
\begin{eqnarray}
\lefteqn{\mathcal{C}_0(a,b,D) = \frac{(D\!-\!2)(D\!+\!2)}{16} + \frac14 \Bigl( 
\frac{D b \!-\! 2}{b \!-\! 2}\Bigr) - \frac1{16} \Bigl( \frac{D b \!-\! 2}{b \!-\! 2}
\Bigr)^2 } \nonumber \\
& & \hspace{5cm} - \frac{(D\!-\!2)(D\!-\!1) a}{8} + \frac{(D \!-\! 2) (D\!-\!1) a}{8 
(b\!-\! 2)^2} \; . \qquad \label{C0D}
\end{eqnarray}

Expression (\ref{simpleM}) can be renormalized by extracting another d'Alembertian
and then adding zero in the form of the equation (\ref{propeqn}) for $i\Delta(x;x')$
\cite{Onemli:2002hr,Onemli:2004mb},
\begin{eqnarray}
\lefteqn{\Bigl[ i\Delta(x;x')\Bigr]^2 = \frac{ \Gamma^2(\frac{D}2 \!-\! 1)}{16 \pi^D} 
\frac{\partial^2}{2 (D \!-\! 3) (D \!-\! 4)} \Bigl[\frac1{\Delta x^{2D-6}} \Bigr] 
\; , } \\
& & \hspace{-.7cm} = \frac{ \Gamma^2(\frac{D}2 \!-\! 1)}{32 (D\!-\!3) (D\!-\! 4) 
\pi^D} \Biggl\{\partial^2 \Bigl[ \frac1{\Delta x^{2D-6}} \!-\! \frac{\mu^{D-4}}{
\Delta x^{D-2}} \Bigr] + \frac{\mu^{D-4} 4 \pi^{\frac{D}2} i \delta^{D}(x \!-\! x')}{
\Gamma(\frac{D}2 \!-\! 1)} \Biggr\} , \qquad \\
& & \hspace{-.7cm} = -\frac{\partial^2}{64 \pi^4} \Bigl[\frac{\ln(\mu^2 \Delta x^2)}{
\Delta x^2} \Bigr] + \frac{\Gamma(\frac{D}2 \!-\! 1) \mu^{D-4} i\delta^D(x \!-\! x')}{
8 (D \!-\! 3)(D\!-\! 4) \pi^{\frac{D}2}} + O(D \!-\! 4) \; . \label{divergence}
\end{eqnarray} 
The second term of (\ref{divergence}) can be absorbed with a local (higher derivative)
counterterm, which gives the renormalized result,
\begin{equation}
-i M^2_{\rm ren}(x;x') = \kappa^2 C_0(a,b) \!\times\! \frac{\partial^6}{64 \pi^4} 
\Bigl[ \frac{\ln(\mu^2 \Delta x^2)}{\Delta x^2} \Bigr] \; . \label{Mren}
\end{equation}
The gauge dependent constant $C_0(a,b)$ is obtained by setting $D=4$ in (\ref{C0D}),
\begin{equation}
C_0(a,b) \equiv \mathcal{C}_0(a,b,4) = \frac34 \frac{(b \!-\! 5) (b \!-\! 1)}{(b 
\!-\! 2)^2} - \frac34 \frac{(b \!-\! 3) (b \!-\! 1) a}{(b \!-\! 2)^2} \; .
\label{C0}
\end{equation}

\subsection{Effective Field Equation} \label{effective}

The scalar self-mass is used to quantum correct its kinetic operator,
\begin{equation}
\partial^2 \phi(x) \longrightarrow \partial^2 \phi(x) - \int \!\! d^4x' \,
M^2_{\rm ren}(x;x') \phi(x') \; .
\end{equation}
We employ the Schwinger-Keldysh formalism \cite{Schwinger:1960qe,Mahanthappa:1962ex,
Bakshi:1962dv,Bakshi:1963bn,Keldysh:1964ud} to obtain real and causal effective 
field equations. There are many good reviews on this subject\cite{Chou:1984es,
Jordan:1986ug,Calzetta:1986ey} so we merely apply the well-known rules for 
converting an in-out result such as (\ref{Mren}) into its Schwinger-Keldysh analog
\cite{Ford:2004wc},
\begin{equation}
M^2_{\rm SK}(x;x') = -\kappa^2 C_0(a,b) \!\times\! \frac{\partial^8}{128 \pi^3}
\Biggl\{ \theta(\Delta t \!-\! \Delta r) \Biggl[ \ln\Bigl[ \mu^2 (\Delta t^2 \!-\!
\Delta r^2)\Bigr] \!-\! 1\Biggr] \Biggr\} . \label{MSK}
\end{equation}
Here $\Delta t \equiv t - t'$ and $\Delta r \equiv \Vert \vec{x} - \vec{x}'\Vert$.

The equation which gives the effective scalar response to a static point source 
of unit strength is,
\begin{equation}
\partial^2 \phi(x) - \int \!\! d^4x' \, M^2_{\rm SK}(x;x') \phi(x') = \delta^3(
\vec{x}) \; . \label{effeqn1}
\end{equation}
Because the four factors of the d`Alermbertian in expression (\ref{MSK}) could be 
considered as acting on either the primed or the un-primed coordinate, we can 
partially integrate one of them and extract the other three from the integration,
\begin{equation}
\partial^2 \phi(x) + \frac{\kappa^2 C_0(a,b) \partial^6}{128 \pi^3} \! \int \!\! d^4x'
\Bigl\{ \qquad \Big\} {\partial'}^2 \phi(x') = \delta^3(\vec{x}) = \partial^2 \Bigl(
-\frac1{4 \pi r}\Bigr) \; , \label{effeqn2}
\end{equation}
where the curly bracketed terms of equations (\ref{MSK}) and (\ref{effeqn2}) are the 
same. Relation (\ref{effeqn2}) is easy to recast as a perturbative solution for 
$\phi(x)$,
\begin{eqnarray}
\phi(x) & = & -\frac1{4\pi r} - \frac{\kappa^2 C_0(a,b) \partial^4}{128 \pi^3} \!
\int_{-\infty}^{t - r} \!\!\!\!\! dt' \Biggl[ \ln\Bigl[\mu^2 (\Delta t^2 \!-\! r^2)
\Bigr] \!-\! 1\Biggr] + O(\kappa^4) \; , \qquad \\
& = & -\frac1{4\pi r} \Biggl\{1 + \frac{\kappa^2 C_0(a,b)}{8 \pi^2 r^2} + O(\kappa^4)
\Biggr\} \; . \label{phicor}
\end{eqnarray}

Relation (\ref{phicor}) purports to be the one loop quantum gravitational correction 
to the long range massless scalar potential induced by a static point source. Much of
the result makes good sense. There should be a quantum gravitational correction to 
this potential because the tree order result distorts virtual gravitons in the vicinity 
of the source. The factional correction of $\kappa^2/r^2$ is dictated by dimensional 
analysis and the single loop counting parameter. However, the overall factor of 
$C_0(a,b)$ is completely unacceptable. By varying the parameters $a$ and $b$ in 
expression (\ref{C0}) we see that $C_0(a,b)$ can be made to range from $-\infty$ to 
$+\infty$!

\section{DeWitt's Lost Theorem} \label{bryce}

This is not a new problem. As discussed in the Introduction, it is usually
resolved by appealing to the S-matrix. However, in 1981 Bryce DeWitt made 
this intriguing statement about choosing different gauge fixing terms for the 
quantum correction $\Sigma$ to the action in the background field formalism
\cite{DeWitt:1981cx}:
\begin{quote}
The functional form of $\Sigma$ is not independent of the choice of these terms.
However, the solutions of the {\it effective field equation}
\begin{eqnarray}
0 = \frac{\delta \Gamma}{\delta g_{\mu\nu}} = \frac{\delta S}{\delta g_{\mu\nu}}
+ \frac{\delta \Sigma}{\delta g_{\mu\nu}} \nonumber 
\end{eqnarray}
can be shown to be the same for all choices.$^{15,16,20}$
\end{quote}
None of the references DeWitt cited\footnote{In order, his references are ${}^{15}$ 
a proceedings article by `t Hooft \cite{tHooft:1975uxh}, ${}^{16}$ another 
proceedings article by DeWitt \cite{DeWitt:1980jv}, and ${}^{20}$ the comment, 
{\it The S-matrix (built out of the tree amplitudes of $\Gamma$) is also choice 
independent.} The comment indicates that DeWitt distinguished between solutions 
and the effective action evaluated at a general solution, which is a generating 
functional for the S-matrix.} provides an explicit proof of this statement but 
we believe he was referring to how one uses asymptotic scattering data to 
parameterize solutions to the effective field equations. 

The background field effective action is gauge invariant, but dependent upon the
gauge which was used to compute quantum corrections to the classical action.
When solving the resulting effective field equations for the metric, and other 
gauge fields, one must of course fix the gauge to get a definite solution, and 
the solution will depend in the usual way on that gauge choice. However, DeWitt 
was discussing the functional dependence upon the quantum gauge fixing term.

To simplify the argument we work in the context of a scalar field $\varphi(x)$
whose renormalized effective action is $\Gamma[\varphi]$. Just being a solution 
of the effective field equation does not eliminate gauge dependence. The key to 
getting a gauge independent result is to correctly normalize the linearized 
solution and then use perturbation theory to expand this into a full solution of 
the effective field equation. If the plane wave mode function for wave vector 
$\vec{k}$ is $u(t,k)$, and the full (gauge dependent) field strength 
renormalization is $Z$, the correct linearized solution is,
\begin{equation}
\varphi_1[\alpha,\alpha^*](x) = \int \!\! \frac{d^3k}{(2\pi)^3} \Biggl\{ 
\frac{u(t,k)}{\sqrt{Z}} e^{i \vec{k} \cdot \vec{x}} \alpha(\vec{k}) +
\frac{u^*(t,k)}{\sqrt{Z}} e^{-i \vec{k} \cdot \vec{x}} \alpha^*(\vec{k}) \Biggr\}
\; . \label{linsol}
\end{equation}
Here the complex parameters $\alpha(\vec{k})$ and $\alpha^*(\vec{k})$ characterize
which of the infinitely many possible linearized solutions is desired.

The effective field equation can be written in terms of a ``scattering current''
which is only nonzero at some early time $t_{\rm in}$ and some late time 
$t_{\rm out}$ \cite{DeWitt:1985bc},
\begin{equation}
J_{\infty}[\alpha,\alpha^*](x) \equiv \varphi_1(x) (\overleftarrow{\partial}_t 
\!-\! \overrightarrow{\partial}_t) \delta(t \!-\! t_{\rm out}) -  \varphi_1(x)
(\overleftarrow{\partial}_t \!-\! \overrightarrow{\partial}_t) \delta(t \!-\! 
t_{\rm in}) \; . \label{Jinfty}
\end{equation}
The role of $J_{\infty}$ is to inject linearized solutions in the asymptotic past
and remove them in the asymptotic future. Applying perturbation theory to the
full effective field equation allows one to develop (\ref{linsol}) into a full
solution,
\begin{equation}
\frac{\delta \Gamma[\varphi]}{\delta \varphi(x)} = -J_{\infty}[\alpha,\alpha^*](x) 
\quad \Longrightarrow \quad \varphi[\alpha,\alpha^*](x) = 
\varphi_1[\alpha,\alpha^*](x) + O\Bigl(\varphi_1^2\Bigr) \; . 
\label{fullsol}
\end{equation}
It is certainly true that evaluating the effective action at this solution gives a
generating functional for the S-matrix, which is independent of the gauge used to
compute quantum corrections to $\Gamma$. DeWitt seems to be claiming that the 
solutions themselves are also independent of this gauge choice. 

We are not sure this claim is true, but the physics of how it would work seems 
clear enough. The point is that simply solving the effective field equation with
a classical source --- as we did in section~\ref{effective} --- is not enough. 
Some physical source must cause any disturbance in the effective field, and some 
physical observer must measure this disturbance. The source and observer both 
interact with quantum gravity and these interactions must be included to produce 
a gauge independent result. Once this dependence is included one can solve a 
modified effective field equation in a way that makes sense even in cosmology.

\section{Including the Observer \& the Source} \label{obssource}

In this section we add a source and observer in the form of a massive scalar
$\psi$ which couples minimally to gravity and to $\phi$,
\begin{equation}
\mathcal{L}_2 = \frac{R \sqrt{-g}}{16 \pi G} -\frac12 \partial_{\mu} \phi 
\partial_{\nu} \phi g^{\mu\nu} \sqrt{-g} -\frac12 \partial_{\mu} \psi
\partial_{\nu} \psi g^{\mu\nu} \sqrt{-g} - \frac12 \Bigl(m^2 \!+\! 
\lambda \phi\Bigr) \psi^2 \sqrt{-g} \; . \label{Lag2}
\end{equation}
From the scattering of two $\psi$'s one can extract a gauge independent 
measure of the one loop quantum gravitational correction to the $\phi$ 
potential. To do this we find the order $\kappa^2 \lambda^2$ contribution to 
the amputated 4-$\psi$ vertex function. Although many diagrams contribute, the 
analysis can be simplified by exploiting Donoghue's crucial insight that only 
very special, nonanalytic terms modify the long range potential (\ref{phicor})
\cite{Donoghue:1993eb,Donoghue:1994dn}. This essentially eliminates the massive 
scalar propagator. Our analysis closely follows Bjerrum-Bohr's computation of 
quantum gravitational corrections to the Coulomb potential 
\cite{BjerrumBohr:2002sx}. We begin by working out how the self-mass 
(\ref{simpleM}) contributes. Next the graviton correlation between the two 
vertices is computed, and we see that it can be regarded as simply changing 
the gauge-dependent constant $C_0(a,b)$ in expression (\ref{phicor}). Then 
each of the remaining diagrams is subjected to a similar analysis, and the 
gauge independence of the final result is manifest by the cancellation of all 
dependence on the parameters $a$ and $b$.

\vskip 2cm

\begin{figure}[ht]
\hspace{1cm} \includegraphics[width=11.0cm,height=0.5cm]{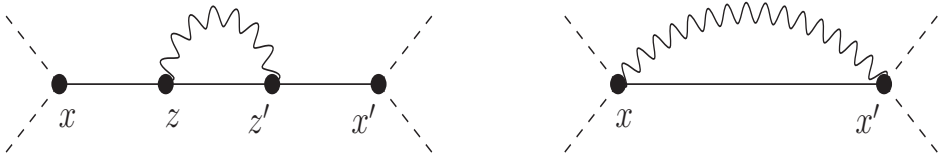}
\caption{The left diagram shows how the self-mass contributes to the
amputated 4-$\psi$ vertex function. The diagram on the right shows how graviton 
correlations between the two vertices contribute. Solid lines represent the 
massless scalar, wavy lines represent the graviton, and dashed lines stand for 
the massive scalar. These graphs have the same topology as Bjerrum-Bohr's 
Diagrams 8 and 4, respectively \cite{BjerrumBohr:2002sx}.}
\label{BB84}
\end{figure}

The massive scalar propagator obeys,
\begin{equation}
(\partial^2 \!-\! m^2) i\Delta_{m}(x;x') = i \delta^D(x \!-\! x') \;\;
\Longrightarrow \;\; i\Delta_m(x;x') = \frac{m^{D-2}}{(2 \pi)^{\frac{D}2}} 
\frac{K_{\frac{D}2 -1}(m \Delta x)}{(m \Delta x)^{\frac{D}2-1}} \; .
\label{massivescalar}
\end{equation}
We do not need it to compute the contribution to the amputated 4-point 
function from the primitive scalar self-mass (\ref{simpleM}). This is the 
leftmost diagram of Figure~\ref{BB84}, which has the same topology as 
Bjerrum-Bohr's Diagram 8 \cite{BjerrumBohr:2002sx}. For our model (\ref{Lag2}) 
we can partially integrate the factors of $\partial^2$ in expression 
(\ref{simpleM}) to eliminate the outer propagators,
\begin{eqnarray}
\lefteqn{-i V_0(x;x') = (-i\lambda) \!\! \int \!\! d^Dz \, i\Delta(x;z) 
\!\times\! (-i\lambda) \!\! \int \!\! d^Dz' \, i\Delta(x';z') \!\times\!
-i M^2(z;z') \; , } \\
& & \hspace{0cm} = \kappa^2 \lambda^2 \mathcal{C}_0(a,b,D) \!\! \int \!\!
d^Dz\, \partial^2_{z} i\Delta(x;z) \!\! \int \!\! d^Dz' \, \partial^2_{z'}
i\Delta(x';z') \!\times\! \Bigl[ i\Delta(z;z')\Bigr]^2 \; , \qquad \\
& & \hspace{0cm} = -\kappa^2 \lambda^2 \mathcal{C}_0(a,b,D) \!\times\!
\Bigl[ i\Delta(x;x')\Bigr]^2 \; ,
 \label{V0}
\end{eqnarray}
where $\mathcal{C}_0(a,b,D)$ is given in~(\ref{C0D}) and  
the contribution to the renormalised self-mass is characterised by 
$C_0(a,b)=\mathcal{C}_0(a,b,4)$ in~(\ref{C0}).

\subsection{Correlations between the Vertices} \label{vertices}

The simplest extra contribution to the amputated 4-point function is the 
rightmost of the diagrams on Figure~\ref{BB84}, which has the same topology
as Bjerrum-Bohr's Diagram 4 \cite{BjerrumBohr:2002sx}. This diagram represents 
quantum gravitational correlations between the source and observer vertices, 
\begin{eqnarray}
-iV_1(x;x') & = & -\frac{i\kappa \lambda}{2} \eta^{\mu\nu} \!\times\! 
i \Bigl[\mbox{}_{\mu\nu} \Delta_{\rho\sigma}\Bigr](x;x') \!\times\! 
-\frac{i\kappa \lambda}{2} \eta^{\rho\sigma} \!\times\! i\Delta(x;x') \; , \\
& = & -\kappa^2 \lambda^2 \biggl[-\frac{2 (D\!-\!1)}{(D\!-\!2) (b\!-\! 2)^2} 
\!+\! \frac{a}{(b \!-\! 2)^2} \biggr] \!\times\! \Bigl[ i\Delta(x;x')\Bigr]^2
\; . \label{V1} 
\end{eqnarray}
Comparing expressions (\ref{V0}) and (\ref{V1}) reveals that we can think
of $-iV_1(x;x')$ as a sort of contribution to the self-mass,
\begin{equation}
-i M^2_1(x;x) = -\kappa^2 \mathcal{C}_1(a,b,D) \times \partial^4 
\Bigl[i\Delta(x;x')\Bigr]^2 \; , \label{DM1}
\end{equation}
where the gauge-dependent factor is,
\begin{equation}
\mathcal{C}_1(a,b,D) = -\frac{2 (D\!-\!1)}{(D\!-\!2) (b\!-\! 2)^2} 
\!+\! \frac{a}{(b \!-\! 2)^2} \quad , \quad C_1(a,b) = 
\frac{(a \!-\! 3)}{(b \!-\! 2)^2}  \; .
\end{equation}
Because (\ref{DM1}) takes the same form as (\ref{simpleM}), with the
replacement of $\mathcal{C}_0(a,b,D)$ by $\mathcal{C}_1(a,b,D)$, we just
add $C_0(a,b)$ and $C_1(a,b)$ in expression (\ref{phicor}). 

\subsection{Vertex-Source and Vertex-Observer Correlations} \label{vertsource}

\begin{figure}[ht]
\hspace{1cm} \includegraphics[width=11.0cm,height=6cm]{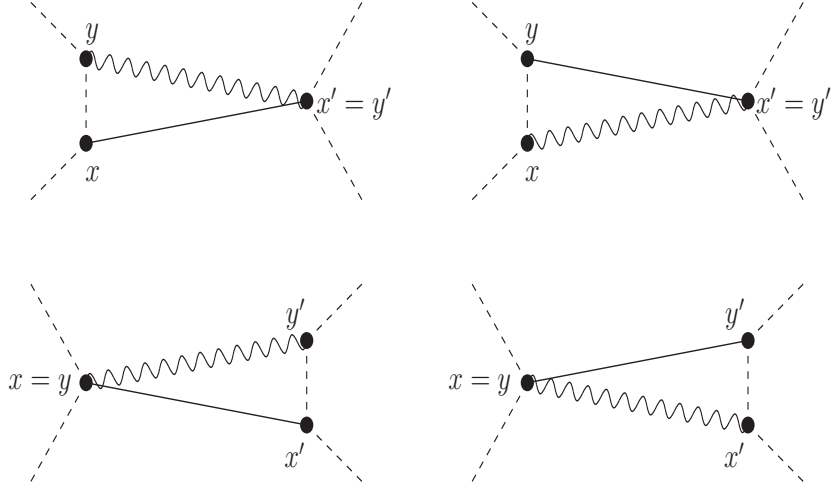}
\caption{These diagrams show correlations between the source (primed) or 
observer (unprimed) and the opposite vertex. Solid lines represent the massless
scalar, wavy lines represent the graviton, and dashed lines represent the massive
scalar. These graphs have the same topology as Bjerrum-Bohr's Diagram 3
\cite{BjerrumBohr:2002sx}.}
\label{BB3}
\end{figure}

The remaining diagrams involve three or four distinct points. We shall reserve
$x^{\mu}$ and $y^{\mu}$ for the in-coming and out-going observer, respectively,
with ${x'}^{\mu}$ and ${y'}^{\mu}$ for the in-coming and out-going source. This
section concerns the four diagrams of Figure~\ref{BB3}, which have the same 
topology as Bjerrum-Bohr's Diagram 3 \cite{BjerrumBohr:2002sx}. For us these
diagrams represent correlations between the source or observer and the more
distant vertex. Correlations with the nearer vertex are cancelled by field 
strength renormalization and do not contribute to the long range potential
(\ref{phicor}). 
 
The full contribution from these diagrams is,
\begin{eqnarray}
\lefteqn{ -i V_2(x;y;x';y') = (-i\lambda) i\Delta(x;x') (-i\kappa)\Bigl[
-\overline{\partial}^{\mu}_{y} \partial^{\nu}_{y} \!+\! \frac{\eta^{\mu\nu}}{2}
(\overline{\partial}_{y} \!\cdot\! \partial_{y} \!+\! m^2) \Bigr] i\Delta_m(x;y) } 
\nonumber \\
& & \hspace{.5cm} \times i\Bigl[\mbox{}_{\mu\nu} \Delta_{\rho\sigma}\Bigr](y;x')
\!\times\! \Bigl( -\frac{i}{2} \kappa \lambda \eta^{\rho\sigma}\Bigr) \!\times\!
\delta^D(x' \!-\! y') + \Bigl({\rm 3\ permutations}\Bigr) \; , \qquad 
\label{V2}
\end{eqnarray}
where ``({\rm 3\ permutations})" indicates the other 3 diagrams of 
Figure~\ref{BB3}.
Here and henceforth an over-lined derivative indicates that it acts on the
external state. For example, $\overline{\partial}_{y}^{\mu}$ means that the
derivative acts on the out-going observer wave function.

Performing all the contractions and acting all the derivatives in expression 
(\ref{V2}) is quite tedious. It is also unnecessary if one only wants terms
that can contribute to the long range potential (\ref{phicor}), which are
equivalent to the form,
\begin{equation}
-i V_2(x;y;x';y') \longrightarrow -\kappa^2 \lambda^2 \mathcal{C}_2(a,b,D)
\!\times\! \Bigl[i\Delta(x;x')\Bigr]^2 \!\times\! \delta^D(x \!-\! y) 
\delta^D(x' \!-\! y') \; . \label{V2simp} 
\end{equation}
For the purpose of identifying those terms which contribute to the long range 
potential one can make the simplification,
\begin{equation}
i\Delta_m(x;y) i\Delta(x;x') i\Delta(y;x') \longrightarrow 
\frac{i\delta^D(x \!-\! y)}{2 m^2} \Bigl[ i\Delta(x;x')\Bigr]^2 \; . 
\label{IntID1}
\end{equation}
Relation (\ref{IntID1}) is the position-space version of Bjerrum-Bohr's 
equation (B4), with a classical general relativistic contribution dropped
\cite{BjerrumBohr:2002sx}. This relation was originally derived by Donoghue
\cite{Donoghue:1993eb,Donoghue:1994dn}. When it is used, along with the
the propagator equations and the fact that $\overline{\partial}^2_{y} - m^2 
= 0$, the result is surprising,
\begin{equation}
C_2(a,b) = \mathcal{C}_2(a,b,4) = 0 \; .
\label{C 2}
\end{equation}
Some details of the derivation of~(\ref{C 2}) are quite technical and are 
therefore given in the Appendix.

\subsection{Vertex-Force Carrier Correlations} \label{vertforce}

\vskip 1cm

\begin{figure}[ht]
\includegraphics[width=13.0cm,height=1cm]{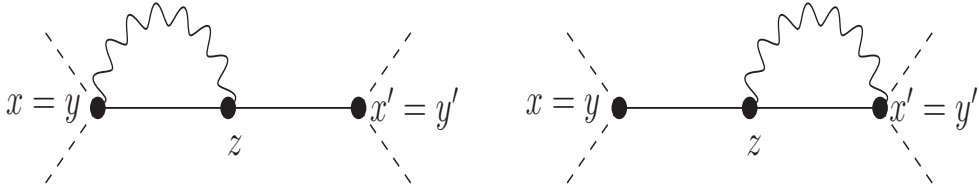}
\caption{These diagrams show correlations between one of the vertices and
the massless scalar force carrier. Solid lines represent the massless 
scalar, wavy lines represent the graviton, and dashed lines stand for the 
massive scalar. These graphs have the same topology as Bjerrum-Bohr's Diagram 7
\cite{BjerrumBohr:2002sx}.}
\label{BB7}
\end{figure}

The massless scalar whose exchange carries the force between source and observer
also interacts with gravity, so we must include quantum gravitational correlations 
between it and the vertices. The relevant graphs are shown in Figure~\ref{BB7},
and they have the same topology as Bjerrum-Bohr's Diagram 7 \cite{BjerrumBohr:2002sx}.
The contribution they make to the amputated 4-$\psi$ vertex function is,
\begin{eqnarray}
\lefteqn{-i V_3(x;y;x';y') = \delta^D(x \!-\! y) \delta^D(x' \!-\! y') \!\int \!\!
d^Dz \Bigl(-\frac{i}{2} \kappa\lambda \eta^{\mu\nu}\Bigr) i\Bigl[\mbox{}_{\mu\nu}
\Delta_{\rho\sigma}\Bigr](x;z) } \nonumber \\
& & \hspace{-.5cm} \times i\Delta(x;z) (-i\kappa) 
\Bigl[-\overleftarrow{\partial}^{\rho}_{z} \overrightarrow{\partial}^{\sigma}_{z}
\!+\! \frac{\eta^{\rho\sigma}}{2} \overleftarrow{\partial}_z \!\!\cdot\!\! 
\overrightarrow{\partial}_z \Bigr] i\Delta(z;x') (-i\lambda) \!+\! \Bigl({\rm 
permutation}\Bigr) . \qquad \label{V3}
\end{eqnarray}

The structure in expression (\ref{V3}) is simple enough that we can explain its
reduction in detail. One key point is to evaluate the scalar derivatives on the 
second line,
\begin{eqnarray}
\lefteqn{ i\Delta(x;z) \Bigl[-\overleftarrow{\partial}^{\rho}_{z} \overrightarrow{\partial}^{\sigma}_{z} \!+\! \frac{\eta^{\rho\sigma}}{2} 
\overleftarrow{\partial}_z \!\!\cdot\!\! \overrightarrow{\partial}_z \Bigr] 
i\Delta(z;x') } \nonumber \\
& & \hspace{2cm} = \frac{(D \!-\! 2) i\Delta(x;z)}{(x \!-\! z)^2} \Bigl[-
(x \!-\! z)^{\rho} \eta^{\sigma\alpha} \!+\! \frac{\eta^{\rho\sigma}}{2}
(x \!-\! z)^{\alpha} \Bigr] \frac{\partial i\Delta(z;x')}{\partial z^{\alpha}}
. \quad \label{red1} 
\end{eqnarray}
A second key point is the expressing the contracted graviton propagator in terms
of the massless scalar propagator,
\begin{eqnarray}
\lefteqn{\eta^{\mu\nu} i\Bigl[\mbox{}_{\mu\nu} \Delta_{\rho\sigma}\Bigr](x;z) 
= \Biggl\{ \frac{4 \eta_{\rho\sigma}}{(D \!-\! 2) (b \!-\! 2)} } \nonumber \\
& & \hspace{1cm} + \frac{2 [(D \!-\! 2) a \!-\! D b \!+\! 2]}{(D \!-\! 2) 
(b \!-\! 2)^2} \Biggl[\eta_{\rho\sigma} - \frac{(D \!-\! 2) (x \!-\! z)_{\rho} 
(x \!-\! z)_{\sigma}}{(x \!-\! z)^2} \Biggr] \Biggr\} i\Delta(x;z) \; . \quad 
\label{red2} 
\end{eqnarray}
Contracting (\ref{red2}) into (\ref{red1}) can be expressed as a derivative of
the square of $i\Delta(x;z)$,
\begin{eqnarray}
\lefteqn{ \eta^{\mu\nu} i\Bigl[\mbox{}_{\mu\nu} \Delta_{\rho\sigma}\Bigr](x;z)
\!\times\! i\Delta(x;z) \Bigl[-\overleftarrow{\partial}^{\rho}_{z} 
\overrightarrow{\partial}^{\sigma}_{z} \!+\! \frac{\eta^{\rho\sigma}}{2} 
\overleftarrow{\partial}_z \!\!\cdot\!\! \overrightarrow{\partial}_z \Bigr] 
i\Delta(z;x') = \Biggl[ \frac{2 (D\!-\! 2)}{(b \!-\! 2)} } \nonumber \\
& & \hspace{0.5cm} + \frac{2 (D \!-\! 2) [(D \!-\! 2) a \!-\! D b \!+\! 2]}{
(b \!-\! 2)^2} \Biggr] \!\times\! \frac{ (x \!-\! z)^{\alpha}}{(x \!-\! z)^2} 
\Bigl[ i\Delta(x;z)\Bigr]^2 \!\times\! \frac{\partial i\Delta(z;x')}{\partial 
z^{\alpha}} \; , \qquad \\
& & \hspace{-0.7cm} = \Biggl[ -\Bigl( \frac{D \!-\! 1}{b \!-\! 2}\Bigr) \!+\! 
\frac{(D \!-\! 2) a \!-\! 2 (D \!-\! 1)}{(b \!-\! 2)^2} \Biggr] \!\times\!
\eta^{\alpha\beta} \frac{\partial}{\partial z^{\alpha}} \Bigl[ i\Delta(x;z)
\Bigr]^2 \!\times\! \frac{\partial}{\partial z^{\beta}} i\Delta(z;x') \; .
\quad \label{red3}
\end{eqnarray}
Of course the next step is to partially integrate on $z$ and use the propagator
identity,
\begin{equation}
-\eta^{\alpha\beta} \frac{\partial}{\partial z^{\alpha}} \frac{\partial}{\partial
z^{\beta}} i\Delta(z;x') = -i\delta^D(z \!-\! x') \; .
\end{equation}
The same reductions apply to the other diagram and the final result takes the
form,
\begin{equation}
-i V_3(x;y;x';y') = -\kappa^2 \lambda^2 \mathcal{C}_3(a,b,D) \!\times\!
\Bigl[ i\Delta(x;x')\Bigr]^2 \!\times\! \delta^D(x \!-\! y) \delta^D(x' \!-\! y')
\; , 
\end{equation}
where the gauge dependent multiplicative factor is,
\begin{equation}
\mathcal{C}_3(a,b,D) = \Bigl(\frac{D \!-\!1}{b \!-\! 2}\Bigr) - 
\frac{[(D \!-\! 2) a - 2 (D\!-\!1)]}{(b \!-\! 2)^2} \; \Longrightarrow \;
C_3(a,b) = \frac{3}{b \!-\! 2} - \frac{2 (a \!-\! 3)}{(b \!-\! 2)^2} .
\end{equation}

\subsection{Source-Observer Correlations} \label{sourceobs}

\begin{figure}[ht]
\hspace{1cm} \includegraphics[width=11.0cm,height=6cm]{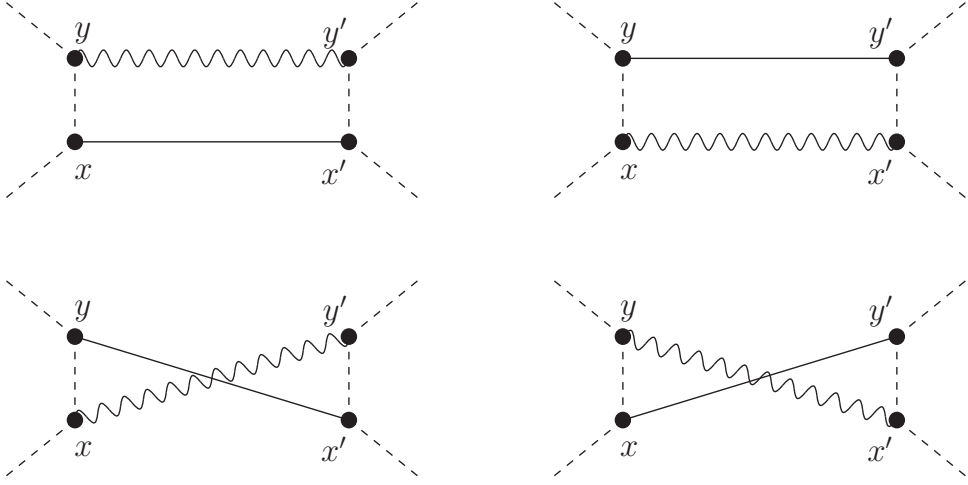}
\caption{These diagrams show correlations between the source (primed) and the
observer (unprimed). Solid lines represent the massless scalar, wavy lines 
represent the graviton, and dashed lines represent the massive scalar. These 
graphs have the same topology as Bjerrum-Bohr's Diagram 2
\cite{BjerrumBohr:2002sx}.}
\label{BB2}
\end{figure}

Both the source and the observer interact with quantum gravity so we must
include graviton correlations between them. The four graphs which contribute
to the long range potential are shown in Figure~\ref{BB2}. (Correlations
from source to source or observer to observer do not affect long range
potential.) The exact contribution for these diagrams is,
\begin{eqnarray}
\lefteqn{ -i V_4(x;y;x';y') = (-i\lambda)^2 i\Delta(x;x') (-i\kappa) 
\Bigl[-\overline{\partial}^{\mu}_{y} \partial^{\nu}_{y} \!+\! 
\frac{\eta^{\mu\nu}}{2} (\overline{\partial}_{y} \!\cdot\! \partial_{y} 
\!+\! m^2)\Bigr] i\Delta_m(y;x) } \nonumber \\
& & \hspace{-0.5cm} \times i\Bigl[\mbox{}_{\mu\nu} \Delta_{\rho\sigma}
\Bigr](y;y') (-i\kappa) \Bigl[-\overline{\partial}^{\rho}_{y'} 
\partial^{\sigma}_{y'} \!+\! \frac{\eta^{\rho\sigma}}{2} 
(\overline{\partial}_{y'} \!\!\cdot\! \partial_{y'} \!+\! m^2)\Bigr] 
i\Delta_m(y';x') \!+\! \Bigl({\rm 3\ P's}\Bigr) . \qquad \label{V4}
\end{eqnarray}
Here ``(3 P's)'' stands for the other three diagrams of Figure~\ref{BB2},
which are simple permutations of the expression shown. We also recall that 
an over-lined derivative indicates it acting on the appropriate external 
state wave function.

The reduction of these diagrams proceeds according to the same methods
as before. In addition to the propagator equations and the simplification
(\ref{IntID1}) the long range potential (\ref{phicor}) is not affected
by the following simplifications,
\begin{eqnarray}
\lefteqn{m^2 (\partial_x \!+\! \partial_y)^{\alpha} (\partial_x \!+\! 
\partial_y)_{\alpha} \Bigl[ i\Delta(x;x') i\Delta(y;y') i\Delta_m(x;y)
i\Delta_m(x';y')\Bigr] } \nonumber \\
& & \hspace{5cm} \longrightarrow -\Bigl[ i\Delta(x;x')\Bigr]^2 
\delta^D(x \!-\! y) \delta^D(x' \!-\! y') \; , \qquad \label{IntID2} \\
\lefteqn{m^2 (\partial_x \!+\! \partial_y)^{\alpha} (\partial_x \!+\! 
\partial_y)_{\alpha} \Bigl[ i\Delta(x;y') i\Delta(y;x') i\Delta_m(x;y)
i\Delta_m(x';y')\Bigr] } \nonumber \\
& & \hspace{5cm} \longrightarrow +\Bigl[ i\Delta(x;x')\Bigr]^2 
\delta^D(x \!-\! y) \delta^D(x' \!-\! y') \; , \qquad \label{IntID3}
\end{eqnarray}
Relations (\ref{IntID2}) and (\ref{IntID3}) are position-space versions of
Bjerrum-Bohr's equations (B8) and (B9), respectively, with some classical
general relativistic contributions neglected \cite{BjerrumBohr:2002sx}.
Both relations were originally derived by Donoghue and Torma 
\cite{Donoghue:1996mt}.

Because $-iV_4(x;y;x';y')$ is ultraviolet finite we can take $D=4$. The final
result for the part relevant to the long range potential (\ref{phicor}) takes
the form,
\begin{equation}
-i V_4(x;y;x';y') \longrightarrow -\kappa^2 \lambda^2 C_4(a,b) \!\times\!
\Bigl[ i\Delta(x;x')\Bigr]^2 \!\times\! \delta^D(x \!-\! y) \delta^D(x' \!-\!
y') \; .
\end{equation}
The gauge-dependent multiplicative factor is,
\begin{equation}
C_4(a,b) = \frac{17}4 - \frac34a - \frac14 \frac{(a \!-\! 3)}{
(b \!-\! 2)^2} \; . 
\end{equation}

\subsection{Force Carrier Correlations with Source \& Observer} \label{forcesource}

\begin{figure}[ht]
\hspace{1cm} \includegraphics[width=11.0cm,height=6cm]{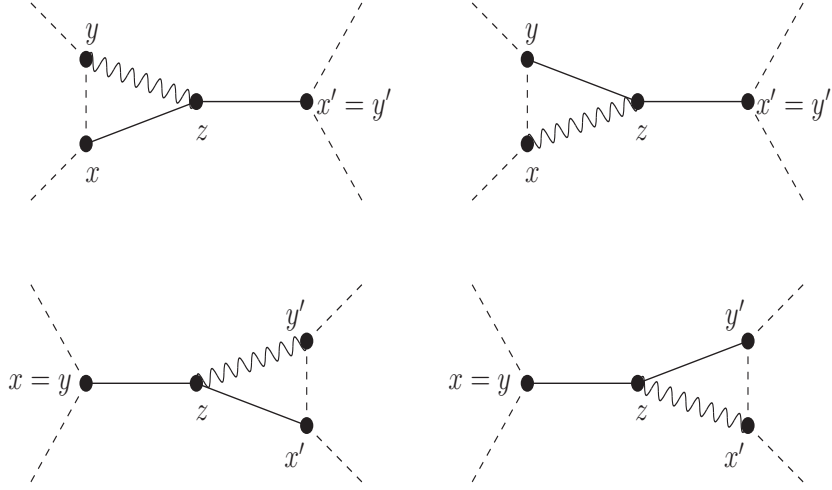}
\caption{These diagrams show correlations between the source (primed) or 
observer (unprimed) and the massless scalar force carrier. Solid lines represent 
the massless scalar, wavy lines represent the graviton, and dashed lines represent 
the massive scalar. These graphs have the same topology as Bjerrum-Bohr's Diagram 
6 \cite{BjerrumBohr:2002sx}.}
\label{BB6}
\end{figure}

Because the source, observer, and the massless scalar which carries the force between
them, all interact with gravity we must include quantum gravitational correlations 
between them. The relevant Feynman diagrams are shown in Figure~\ref{BB6}. Their full
contribution to the amputated 4-$\psi$ vertex function is,
\begin{eqnarray}
\lefteqn{ -i V_5(x;y;x';y') = \kappa^2 \lambda^2 \delta^D(x' \!-\! y')\Bigl[ 
-\overline{\partial}^{\mu}_{y} \partial^{\nu}_{y} \!+\! \frac{\eta^{\mu\nu}}{2}
(\overline{\partial}_{y} \!\cdot\! \partial_{y} \!+\! m^2)\Bigr] i\Delta_m(y;x) 
\!\! \int \!\! d^Dz} \nonumber \\
& & \hspace{0cm} \times i\Bigl[\mbox{}_{\mu\nu} \Delta_{\rho\sigma}\Bigr](y;z) 
\!\times\! i\Delta(x;z) \Bigl[-\overleftarrow{\partial}^{\rho}_{z}
\overrightarrow{\partial}^{\sigma}_{z} \!+\! \frac{\eta^{\rho\sigma}}{2} 
\overleftarrow{\partial}_z \!\!\cdot\!\! \overrightarrow{\partial}_z \Bigr] 
i\Delta(z;x') \!+\! \Bigl({\rm 3\ P's}\Bigr) . \qquad \label{V5}
\end{eqnarray}
As before, the symbol ``(3 P's)'' refers to the three other diagrams shown in 
Figure~\ref{BB6}. Also as before, the over-lined derivative $\overline{\partial}_{y}$
acts on the out-going observer's external wave function.

The reduction of (\ref{V5}) follows previous reductions:
\begin{itemize}
\item{The derivatives with respect to $y$ on the first line of (\ref{V5}) are 
treated the same way as those of expression (\ref{V2}). They are first expressed 
as $m^2$ plus a sum of distinct kinetic operators, then acted on the propagators.
Finally, the simplification (\ref{IntID1}) is invoked.} 
\item{The derivatives with respect to $z$ on the second line of (\ref{V5}) are 
treated the same way as those of expression (\ref{V3}). We first act them on 
the two massless scalar propagators, then the derivative of $i\Delta(x;z)$ is 
combined with the factor of $i\Delta(x;z)$ in the graviton propagator to give a 
total derivative, which is partially integrated onto $i\Delta(z;x')$ to produce 
a delta function that eliminates the integration over $z$.}
\end{itemize}
The final result takes the form,
\begin{equation}
-i V_5(x;y;x';y') = -\kappa^2 \lambda^2 C_5(a,b) \!\times\! \Bigl[ i\Delta(x;x')
\Bigr]^2 \!\times\! \delta^D(x \!-\! y) \delta^D(x' \!-\! y') \; .
\end{equation}
The gauge dependent multiplicative factor is,
\begin{equation}
C_5(a,b) = -2 + \frac32 a - \frac{\frac32}{b \!-\! 2} + \frac12 \frac{(a \!-\! 3)}{
(b \!-\! 2)^2} \; .
\end{equation} 

\subsection{Sum Total} \label{sum}

\begin{table}
\setlength{\tabcolsep}{8pt}
\def\arraystretch{1.5}
\centering
\begin{tabular}{|@{\hskip 1mm }c@{\hskip 1mm }||c|c|c|c|}
\hline
$i$ & $1$ & $a$ & $\frac1{b-2}$ & $\frac{(a-3)}{(b-2)^2}$ \\
\hline\hline
0 & $+\frac34$ & $-\frac34$ & $-\frac32$ & $+\frac34$ \\
\hline
1 & $0$ & $0$ & $0$ & $+1$ \\
\hline
2 & $0$ & $0$ & $0$ & $0$ \\
\hline
3 & $0$ & $0$ & $+3$ & $-2$ \\
\hline
4 & $+\frac{17}4$ & $-\frac34$ & $0$ & $-\frac14$ \\
\hline
5 & $-2$ & $+\frac32$ & $-\frac32$ & $+\frac12$ \\
\hline\hline
Total & $+3$ & $0$ & $0$ & $0$ \\
\hline
\end{tabular}
\caption{The gauge dependent factors $C_i(a,b)$ for each contribution, where
the index $i$ ($i = 1,\dots,5$) refers to the diagrams shown in Figure~$i$.}
\label{Cab}
\end{table}

No other diagrams contribute to the long range potential 
(\ref{phicor}).\footnote{In the SQED computation of Bjerrum-Bohr there was
an additional contribution from what he termed Diagram 5 
\cite{BjerrumBohr:2002sx}. However, this diagram happens to vanish for
the massless scalar process we consider.}
As we have seen, each diagram could be viewed as making a contribution to
the self-mass of the form,
\begin{equation}
-i M^2_i(x;x') = -\kappa^2 C_i(a,b) \!\times\! \partial^4 \Bigl[ 
i\Delta(x;x')\Bigr]^2 \; .
\end{equation}
Table~\ref{Cab} gives our results for the gauge-dependent multiplicative
factors $C_i(a,b)$. It is reassuring that all dependence on the gauge 
parameters $a$ and $b$ drops out in the sum, $\sum_{i=1}^5 C_i(a,b)=+3$. 

The simplest gauge is obtained by setting $a = b = 1$. This was the choice
made by Donoghue \cite{Donoghue:1993eb,Donoghue:1994dn}, and by Bjerrum-Bohr
\cite{BjerrumBohr:2002sx}. It is amusing to note that the actual self-mass,
$-iM_0(x;x')$, we computed in~(\ref{simpleM}) to motivate the problem, 
happens to vanish in that gauge. Of course our final result is independent 
of $a$ and $b$, as would be those of Donoghue and Bjerrum-Bohr had they
made their computations in a general gauge.

\section{Discussion} \label{discuss}

The continual creation of horizon-scale gravitons during inflation tends
to engender secular corrections to particle kinematics \cite{Miao:2005am,
Miao:2006gj,Miao:2007az,Miao:2012bj,Leonard:2013xsa,Wang:2014tza,
Glavan:2015ura,Glavan:2016bvp,Tsamis:1996qk,Mora:2013ypa,Boran:2014xpa,
Boran:2017fsx,Boran:2017cfj} and to force laws \cite{Glavan:2013jca}. It has
even been proposed that the self-gravitation between these gravitons induces
a secular slowing of the expansion rate as more and more of them come into 
causal contact \cite{Tsamis:1996qq,Tsamis:1996qm}. However, behind all of 
these effects lurks the {\it gauge issue}: the simplest way to study what
inflationary gravitons do is from solutions to the effective field 
equations and those solutions depend upon how the graviton's gauge freedom 
is fixed. Some researchers dismiss gauge-dependent Green's functions as 
completely unphysical \cite{Garriga:2007zk}. Others reflect that even 
gauge-dependent Green's functions must contain physical information because 
the flat space S-matrix --- which is gauge independent --- is formed by 
taking sums of products of them \cite{Tsamis:2008zz}. The question is how to 
separate the physical information from the rest.

Our goal has been to develop an analog of the S-matrix which does not
involve the global integrations that preclude the S-matrix from being 
observable in cosmology. We believe the gauge dependence of solutions 
to the usual effective field equations derives from neglecting quantum 
gravitational correlations with the source which disturbs the effective
field and the observer who measures the disturbance. Including these 
correlations leads to an improved effective field equation which can be 
solved quasi-locally. As a test of this idea we worked in the most general
Poincar\'e invariant gauge (\ref{gauge}) to compute the one graviton loop 
correction to the long range force exerted by a massless, minimally coupled 
scalar $\phi$. The conventional result (\ref{phicor}) is highly gauge 
dependent; by varying the two gauge parameters it can be made to go from 
$-\infty$ to $+\infty$! However, including a physical source and observer 
--- in the form of a massive scalar $\psi$ --- led to the complete 
cancellation of gauge dependence which is evident in Table~\ref{Cab}.

Our final result for the effective field equation takes the form,
\begin{equation}
\partial^2 \phi(x) - \int \!\! d^4x' \, M^2_{\rm full}(x;x') \phi(x')
= J(x) \; , \label{finaleqn}
\end{equation}
where the improved, gauge-independent scalar self-mass is,
\begin{equation}
M^2_{\rm full}(x;x') = -\frac{3 \kappa^2 \partial^8}{128 \pi^3} \Biggl\{
\theta(\Delta t \!-\! \Delta r) \Biggl[ \ln\Bigl[ \mu^2 (\Delta t^2 \!-\!
\Delta r^2)\Bigr] - 1\Biggr] \Biggr\} + O(\kappa^4) \; ,
\end{equation}
and we recall that $\Delta t \equiv t - t'$ and $\Delta r \equiv \Vert \vec{x}
- \vec{x}'\Vert$. We actually only considered $J(t,\vec{x}) = 
\delta^3(\vec{x})$ but the equation is linear, so the passage to general 
$J(x)$ follows from superposition. Note that there is no dependence on
the $\psi$ mass $m$. It dropped out through using relations (\ref{IntID1})
and (\ref{IntID2}-\ref{IntID3}) to extract the special nonanalytic part 
of the general amplitude which contributes to the long range potential 
(\ref{phicor}). Realizing that quantum gravitational corrections to low
energy physics derive solely from these special sorts of terms was Donoghue's
great contribution \cite{Donoghue:1993eb,Donoghue:1994dn}; we have merely
translated his relations to position space. We should also comment that
we were greatly aided in recognizing the handful of relevant diagrams by 
the computation Bjerrum-Bohr made of the quantum gravitational correction 
to the Coulomb potential in SQED \cite{BjerrumBohr:2002ks}.

The point of this exercise was to use the flat space S-matrix to abstract
observables for cosmology. The essentials of how to study changes in kinematics
and force laws seem clear enough now:
\begin{itemize}
\item{We want to correct the linearized effective field equation in position 
space;}
\item{We need to include quantum gravitational correlations with the source
which disturbs the effective field, and with the observer who measures the
disturbance; and}
\item{Most details of the source and observer will drop out in the appropriate
infrared limit.}
\end{itemize}
What is not yet apparent is the correct generalization of the relations 
(\ref{IntID1}) and (\ref{IntID2}-\ref{IntID3}) which were used to 
extract the essential part of the full amplitude. In flat space background 
the appropriate infrared limit is large distances. For inflationary cosmology 
we suspect it is late times.

One crucial point which we have not addressed is what observables stand
for the primordial power spectra when one includes loop corrections. The
naive correlators cannot be right because they depend upon the infrared
cutoffs which must be introduced to define the scalar and graviton propagators
\cite{Giddings:2010nc}. Nonlocal composite operator generalizations can be
devised which avoid this dependence \cite{Urakawa:2010it,Gerstenlauer:2011ti}
but these generalizations introduce new ultraviolet divergences and also
disrupt the careful pattern by which loop corrections to the naive 
correlators are slow roll suppressed \cite{Miao:2012xc}.

Finally, we should comment on the other alternative for extracting 
cosmological observables: taking expectation values of gauge 
invariant operators. A recent computation on de Sitter background shows 
how this can be done to invariantly quantify the back-reaction on inflation 
\cite{Miao:2017vly}. To invariantly study changes in particle kinematics,
and the associated force laws, one might compute 2-point functions at 
geodesically fixed separations, as Fr\"ob has recently done for a massless 
scalar on flat space background \cite{Frob:2017apy}. Fr\"ob's calculation 
is interesting because the double logarithms he found would translate to 
corrections to the potential of the form $-\frac1{4\pi r} \times 
\frac{\kappa^2}{r^2} \times \ln(\mu r)$. (The same thing is bound to happen 
for the pure gravitational analog \cite{Tsamis:1989yu}.) Of course the
S-matrix technique we have pursued lacks the factor of $\ln(\mu r)$, so
there is a clear difference between two, completely gauge independent
results. It is not that one is right and the other wrong; they both
represent correct answers to different questions. There is simply no
alternative to thinking hard about how an effect is measured, and then
correctly modelling that process.

\vskip 1cm

\centerline{\bf Acknowledgements}

We are grateful for D. Glavan for locating an old proceedings article
\cite{tHooft:1975uxh}. We are also grateful for correspondence and 
conversation on this subject with S. Deser, J. F. Donoghue, M. B. Fr\"ob 
and G. `t Hooft. This work was partially supported by Taiwan MOST grant 
103-2112-M-006-001-MY3; by the D-ITP consortium, a program of the 
Netherlands Organization for Scientific Research (NWO) that is funded by 
the Dutch Ministry of Education, Culture and Science (OCW); by NSF grant 
PHY-1506513; and by the Institute for Fundamental Theory at the University 
of Florida.

\section*{Appendix: Reduction of the diagrams in Fig.~\ref{BB3}}

In this  Appendix we present some details of the evaluation of the 4-$\psi$ 
vertex~(\ref{V2}). Upon partial integration of $\bar\partial_y$ derivatives, 
$-i V_2(x;y;x';y')$ becomes,
\begin{eqnarray}
-i V_2(x;y;x';y') \!&=&\! \frac{i\lambda^2\kappa^2}{2}\delta^D(x' \!-\! y')  
i\Delta(x;x') \label{Appendix:V2} \\
& & \hspace{-.7cm}
\times\,\,\bigg\{\partial^{\alpha}_{y}\bigg[\Bigl(
\delta^\mu_\alpha\partial^\nu_y\!-\! \frac{1}{2}\eta^{\mu\nu}
\partial^{y}_\alpha
\Bigr) i\Delta_m(x;y)\eta^{\rho\sigma}  i\Bigl[\mbox{}_{\mu\nu} \Delta_{\rho\sigma}\Bigr](y;x')\bigg]
\nonumber \\
& & \hspace{.cm}
+\frac{m^2}{2} i\Delta_m(x;y)\eta^{\mu\nu}\eta^{\rho\sigma} 
 i\Bigl[\mbox{}_{\mu\nu} \Delta_{\rho\sigma}\Bigr](y;x')\bigg\}
+ \Bigl({\rm 3\ perm's}\Bigr) \; , \quad 
\nonumber
\end{eqnarray}
where
\begin{eqnarray}
\eta^{\rho\sigma} i\Bigl[\mbox{}_{\mu\nu} \Delta_{\rho\sigma}\Bigr](y;x')
\!&=&\!\frac{4\eta_{\mu\nu}}{(D\!-\!2)(b\!-\!2)} \, i\Delta(y;x')
\label{Appendix: contraction graviton}\\
\!&&\!\hskip -1.4cm
+\,4\bigg[\frac{a}{(b\!-\!2)^2}
                    -\frac{D}{(D\!-\!2)(b\!-\!2)}-\frac{2(D\!-\!1)}{(D\!-\!2)(b\!-\!2)^2}\bigg]
                \frac{\partial^y_\mu\partial^y_\nu}{\partial^2_y}i\Delta(y;x')
\,
\nonumber
\end{eqnarray}
and
\begin{eqnarray}
\eta^{\mu\nu}\eta^{\rho\sigma} i\Bigl[\mbox{}_{\mu\nu} \Delta_{\rho\sigma}\Bigr](y;x')
=4\bigg[\frac{a}{(b\!-\!2)^2}-\frac{2(D\!-\!1)}{(D\!-\!2)(b\!-\!2)^2}\bigg]
i\Delta(y;x') \,.
\label{Appendix: contraction graviton:2}
\end{eqnarray}
When these are inserted in~(\ref{Appendix:V2}), and one acts the derivatives, 
one obtains,
\begin{eqnarray}
-i V_2(x;y;x';y') \!&=&\! \frac{i\lambda^2\kappa^2}{2}\delta^D(x' \!-\! y')  i\Delta(x;x') 
\label{Appendix:V2:b} \\
& & \hspace{-1.8cm}
\times\,\bigg\{\!\!-\frac{1}{b\!-\!2}\bigg[2 \partial_y^2 
i\Delta_m(x;y) \!\times\! i\Delta(y;x^\prime)
+ 2 \partial_y^\mu i\Delta_m(x;y)\partial^y_\mu i\Delta(y;x^\prime)
\nonumber \\
& & \hspace{.8cm}
-\frac{2D}{D\!-\!2}m^2i\Delta_m(x;y) i\Delta(y;x^\prime)
\bigg]
\nonumber \\
& & \hspace{-3.cm}
%+\bigg[\frac{a}{(b\!-\!2)^2}
 %                   -\frac{D}{(D\!-\!2)(b\!-\!2)}-\frac{2(D\!-\!1)}{(D\!-\!2)(b\!-\!2)^2}\bigg]
+\bigg[\frac{a}{(b\!-\!2)^2}
                    -\frac{Db\!-\!2}{(D\!-\!2)(b\!-\!2)^2}\bigg]
\bigg[4\partial_y^\mu\partial_y^\nu i\Delta_m(x;y) \!\times\!
               \frac{\partial^y_\mu\partial^y_\nu}{\partial_y^2} i\Delta(y;x^\prime)
\nonumber \\
& & \hspace{-2cm}
+2\partial_y^\mu i\Delta_m(x;y)\partial^y_\mu i\Delta(y;x^\prime)
-2(\partial_y^2\!-\!m^2) i\Delta_m(x;y) \!\times\! i\Delta(y;x^\prime)
\bigg]
\bigg\}
\nonumber \\
& & \hspace{-3.cm}
+ \Bigl({\rm 3\ perm's}\Bigr) \; . \quad 
\nonumber
\end{eqnarray}
To reduce this expression further we shall need some identities.
The first useful identity is the Bjerrum-Bohr's identity~(\ref{IntID1}).
The second one can be obtained by noting that, 
\begin{eqnarray}
\phi_m(y)\partial_y^2\big[i\Delta_m(x;y) i\Delta(y;x^\prime)\big]
 \!&=&\! \phi_m(y)m^2i\Delta_m(x;y) i\Delta(y;x^\prime)
\nonumber \\
& & \hspace{-3.cm} =\,\phi_m(y)\Big[m^2i\Delta_m(x;y) i\Delta(y;x^\prime)
     +i\delta^D(x\!-\!y) i\Delta(y;x^\prime)
\nonumber \\
& & \hspace{-3.cm}
 +\, i\Delta_m(x;y) i\delta^D(y\!-\!x^\prime)
 + 2\partial_y^\mu i\Delta_m(x;y)\partial^y_\mu i\Delta(y;x^\prime)\Big]
\label{Appendix: Identity 1}
\,,\quad
\end{eqnarray}
where $\phi_m(y)$ is the external leg field that satisfies, $(\partial_y^2-m^2)\phi_m(y)=0$.
From~(\ref{Appendix: Identity 1}) one immediately obtains, 
\begin{eqnarray}
 2\partial_y^\mu i\Delta_m(x;y)\partial^y_\mu i\Delta(y;x^\prime)
 \!&=&\! - i\Delta_m(x;y) i\delta^D(y\!-\!x^\prime)-i\delta^D(x\!-\!y) i\Delta(y;x^\prime)
\nonumber \\
& & \hspace{-0.6cm}
\longrightarrow\quad -i\delta^D(x\!-\!y) i\Delta(y;x^\prime)
\label{Appendix: Identity 1b}
\,,\quad
\end{eqnarray}
where the last implication selects only the term which contributes to the
long range potential (\ref{phicor}). Analogous (albeit more tedious) manipulations 
yield the following identity,
\begin{eqnarray}
4\partial_y^\mu\partial_y^\nu i\Delta_m(x;y)
               \frac{\partial^y_\mu\partial^y_\nu}{\partial_y^2} i\Delta(y;x^\prime)
\quad \longrightarrow\quad 3i\delta^D(x-y) i\Delta(y;x^\prime)
\label{Appendix: Identity 1b}
\,.\quad
\end{eqnarray}
When these identities are employed in the vertex function~(\ref{Appendix:V2:b}), 
one sees that both the square bracket multiplying the factor $-1/(b-2)$, and 
the square bracket on the last two lines of~(\ref{Appendix:V2:b}), vanish in 
$D=4$, implying that the whole set of diagrams in Figure~\ref{BB3} contributes 
zero to (\ref{phicor}), thus proving~(\ref{C 2}).

\end{document}